\documentclass[modern]{aastex63}
\usepackage{amsmath}
\usepackage{calc}
\clearpage{}%

\clearpage{}%
\DeclareMathOperator*{\argmax}{argmax}
\newcommand{\AR}{active region}
\newcommand{\dem}{$\mathrm{EM}(T)$}
\newcommand{\twait}[1][]{t_{\textup{wait}#1}}

\defcitealias{barnes_understanding_2019}{Paper I}
\defcitealias{viall_survey_2017}{VK17}
\begin{document}

\title{Understanding Heating in Active Region Cores through Machine Learning II. Classifying Observations}
\author[0000-0001-9642-6089]{W. T. Barnes}
\affiliation{National Research Council Postdoctoral Research Associate residing at the Naval Research Laboratory, Washington, D.C. 20375}
\affiliation{Department of Physics \& Astronomy, Rice University, Houston, TX 77005-1827}
\author[0000-0002-3300-6041]{S. J. Bradshaw}
\affiliation{Department of Physics \& Astronomy, Rice University, Houston, TX 77005-1827}
\author[0000-0003-1692-1704]{N. M. Viall}
\affiliation{NASA Goddard Space Flight Center, Greenbelt, MD 20771}
\correspondingauthor{W. T. Barnes}
\email{will.barnes.ctr@nrl.navy.mil}

\received{24 May 2021}
\revised{8 July 2021}
\accepted{14 July 2021}
\submitjournal{The Astrophysical Journal}

\begin{abstract}
Constraining the frequency of energy deposition in magnetically-closed \AR{} cores requires sophisticated hydrodynamic simulations of the coronal plasma and detailed forward modeling of the optically-thin line-of-sight integrated emission.
However, understanding which set of model inputs best matches a set of observations is complicated by the need for any proposed heating model to simultaneously satisfy multiple observable constraints.
In this paper, we train a random forest classification model on a set of forward-modeled observable quantities, namely the emission measure slope, the peak temperature of the emission measure distribution, and the time lag and maximum cross-correlation between multiple pairs of AIA channels.
We then use our trained model to classify the heating frequency in every pixel of \AR{} NOAA 1158 using the observed emission measure slopes, peak temperatures, time lags, and maximum cross-correlations and are able to map the heating frequency across the entire active region.
We find that high-frequency heating dominates in the inner core of the \AR{} while intermediate frequency dominates closer to the periphery of the \AR{}.
Additionally, we assess the importance of each observed quantity in our trained classification model and find that the emission measure slope is the dominant feature in deciding with which heating frequency a given pixel is most consistent.
The technique presented here offers a very promising and widely applicable method for assessing observations in terms of detailed forward models given an arbitrary number of observable constraints.
\end{abstract}

%
\section{Introduction}\label{sec:introduction}

A central problem in the study of the solar corona is whether EUV and soft X-ray observations of \AR s are consistent with the plasma being heated steadily or impulsively.
Observations of hot plasma by the X-Ray Telescope \citep[XRT,][]{golub_x-ray_2007} on the \textit{Hinode} spacecraft \citep{kosugi_hinode_2007} suggest \AR{} cores are heated steadily \citep[e.g.][]{warren_constraints_2011,winebarger_using_2011}.
Alternatively, observations of cooler ($\sim1$ MK) plasma have been shown to be more consistent with impulsive heating where the plasma is allowed to cool significantly between consecutive heating events \citep[e.g][]{winebarger_evolving_2003,mulu-moore_determining_2011,ugarte-urra_investigation_2006,viall_patterns_2011,viall_evidence_2012}.
More recent work \citep{del_zanna_evolution_2015,bradshaw_patterns_2016} suggests that observations of a single \AR{} may be consistent with both steady and impulsive heating, depending on the location within the \AR{}.
Additionally, a given flux tube at a single location in an \AR{} core may undergo a range of heating frequencies \citep[e.g.][]{cargill_active_2014}.
Collectively, these results suggest that \AR s may be heated by a range of frequencies.
Here, we define the heating frequency in terms of the time between consecutive heating events on a given flux tube.

Two often-used diagnostics of the frequency of energy deposition in the coronal plasma are the \deleted{``cool''} emission measure slope and the time lag, the temporal offset which maximizes the cross-correlation between pairs of imaging channels.
The emission measure distribution, $\mathrm{EM}(T)=\int\mathrm{d}h\,n_e^2$, where $n_e$ is the electron density and the integration is taken along the line of sight (LOS), is well-described by the power-law relationship $\textup{EM}(T)\sim T^a$, for $a>0$, over the temperature range $10^{5.5}\lesssim T\lesssim10^{6.5}$ K \citep{jordan_structure_1975,jordan_structure_1976}.
$a$, the emission measure slope in $\log-\log$ space, parameterizes the width of the emission measure distribution and is a commonly used diagnostic for the heating frequency \citep[e.g.][]{tripathi_emission_2011,winebarger_using_2011,warren_constraints_2011,mulu-moore_can_2011,bradshaw_diagnosing_2012,schmelz_cold_2012,reep_diagnosing_2013,del_zanna_evolution_2015}.

The time lag analysis of \citet{viall_evidence_2012} provides an additional diagnostic of the heating frequency. \citet{viall_patterns_2011} showed that, as the plasma cools, the intensity will peak in successively cooler passbands of the Atmospheric Imaging Assembly \citep[AIA,][]{lemen_atmospheric_2012} on the Solar Dynamics Observatory \citep[SDO,][]{pesnell_solar_2012} spacecraft.
The temporal offset which maximizes the cross-correlation between these intensities is a proxy for the cooling time of the plasma between these channels.
The general convention is that the \added{channel whose peak response occurs at a higher temperature (the ``hot'' channel) precedes the channel whose peak response occurs at a lower temperature (the ``cool'' channel)}.
In this case, cooling plasma produces a positive time lag while a negative time lag indicates heating.
However, the multiple temperature peaks in the AIA response functions create a nuance to this general rule.
The 94 and 131 \AA{} channels are doubly peaked in the range of temperatures over which an impulsive heating event may span (see \autoref{fig:aia-response}).
Not knowing \textit{a priori} what physics occurs where, we follow the convention set by \citet{viall_evidence_2012}, where 94 \AA{} is always assumed to be a hot channel, and $8$ MK plasma dominates the emission, and 131 \AA{} is always assumed to be a cool channel where $0.5$ MK plasma dominates the emission.
When these assumptions do not hold, opposite time lag conventions result.
We remove this ambiguity with the other four AIA channels, since they span the temperatures in between. 

Any viable heating model must account for the range of observed emission measure slopes and time lags \citep[\citetalias{viall_survey_2017} hereafter]{viall_survey_2017}.
However, accurately predicting the distributions of these observables for a given heating model is challenging as several factors are likely to impact these diagnostics, including multiple emitting structures along the LOS and non-equilibrium ionization \citep[e.g.][]{barnes_inference_2016}.

In \citet[\citetalias{barnes_understanding_2019} hereafter]{barnes_understanding_2019}, we forward modeled emission from \AR{} NOAA 1158 as observed by the six EUV channels of AIA.
Using a potential field extrapolation combined with $5\times10^3$ separate instances of the Enthalpy-based Thermal Evolution of Loops model \citep[EBTEL,][]{klimchuk_highly_2008,cargill_enthalpy-based_2012,cargill_enthalpy-based_2012-1,barnes_inference_2016}, we predicted time-dependent intensities in each pixel of the \AR{} for a range of nanoflare heating frequencies.
We defined the heating frequency in terms of the dimensionless ratio 
\begin{equation}\label{eq:heating_types}
    \varepsilon = \frac{\langle\twait\rangle}{\tau_{\textup{cool}}} =
    \begin{cases} 
        0.1, &  \text{high frequency},\\
        1, & \text{intermediate frequency}, \\
        5, & \text{low frequency},
     \end{cases}
\end{equation}
where $\tau_{\textup{cool}}$ is the fundamental cooling timescale due to thermal conduction and radiation \citep[see appendix of][]{cargill_active_2014} and $\langle \twait\rangle$ is the average waiting time between consecutive heating events on a given strand.
Specifically, for the shortest loop in our simulated \AR{}, $\approx20$ Mm, the high-, intermediate-, and low-frequency values of $\langle\twait\rangle$ are approximately $90$ s, $930$ s, and $4700$ s, respectively.
For the longest loop, $\approx245$ Mm, the approximate values of $\langle\twait\rangle$ for the three frequencies are $850$ s, $8500$ s, and $42500$ s.

As in \citetalias{barnes_understanding_2019}, we define a \textit{strand} to be a flux tube with the largest possible isothermal cross-section and the fundamental unit of the corona while a \textit{loop} is an observationally-defined feature and an intensity enhancement relative to the surrounding diffuse emission.
Additionally, for each simulated strand, even within a single heating frequency category, there is a spread in the distribution of heating frequencies due to our chosen dependence of the waiting time on the energy of the event.
This parameterization of the heating is discussed in detail in Section 2.3 of \citetalias{barnes_understanding_2019}.

From our predicted intensities, we computed the emission measure slope as well as the time lag and the maximum cross-correlation for all 15 AIA channel pairs.
We found that signatures of the heating frequency persist in both the emission measure slope and the time lag and that, in particular, negative time lags that occur in pairs with the 131 \AA{} channel provide a possible diagnostic for $\ge10$ MK plasma, provided they are co-spatial with positive time lags in the other channel pairs.

While such predicted diagnostics are useful in understanding how observables respond to the frequency of energy deposition, systematically assessing real observations in terms of said model results is nontrivial.
Attempts to tune model parameters to exactly match a single observation (e.g. a light curve from a single pixel) are not likely to generalize well to other data (i.e. ``overfitting'').
Additionally, purely qualitative comparisons between real data and forward models provide no constraint on the observation with respect to the model inputs, regardless of how sophisticated the simulation may be.

Because of the ability to learn non-linear relationships from arbitrary data, machine learning  is an excellent tool for systematically and quantitatively assessing differences between observations and simulations for a range of input parameters.
Machine learning has a variety and growing number of applications in solar physics, including predicting coronal mass ejections \citep[e.g.][]{bobra_predicting_2016}, classifying flare spectra \citep{panos_identifying_2018}, and inverting optically-thick chromospheric lines \citep{osborne_radynversion_2019}.
In particular, \citet{tajfirouze_time-resolved_2016} trained a neural network on $>10^5$ modeled 94 \AA{} and 335 \AA{} AIA light curves simulated using EBTEL for a large parameter space of heating properties.
Using this trained network, they classified a sample of $\sim4000$ s light curves, extracted from \AR s thought to contain ``very hot'' ($\ge6$ MK) plasma, in terms of the model heating input parameters.
They found that these observations were most consistent with many frequent short-duration events drawn from a power-law distribution with index $\alpha=-1.5$.
Combined with predicted observables from sophisticated forward models, systematic comparisons using machine learning methods are well-poised to place strong constraints on heating properties in \AR s.

In this paper, the second in a series concerned with constraining nanoflare heating properties, we train a machine learning classification model, specifically a random forest classifier, to identify the most probable heating frequency in each pixel of \AR{} NOAA 1158 using the predicted emission measure slopes and peak temperatures, time lags, and maximum cross-correlations from \citetalias{barnes_understanding_2019}.
In \autoref{sec:observations}, we describe how the full 12 hours of multi-wavelength AIA observations are processed and how we compute the slope and peak temperature of the emission measure distribution (\autoref{sec:em_slopes}) and time lag and maximum cross-correlation between AIA channels (\autoref{sec:timelags}).
\autoref{sec:compare} describes the random forest classification model as well as the data preparation procedure (\autoref{sec:data-prep}) and \autoref{sec:feature-combos} and \autoref{sec:feature-importance} show the predicted heating frequency in each pixel for several different combinations of features.
In \autoref{sec:discussion} we discuss the results of our classification model and provide some concluding comments in \autoref{sec:conclusions}.
To our knowledge, this paper represents the first attempt to use multiple diagnostics and machine learning to map the heating properties across an observed \AR{}. %
\section{Observations and Analysis}\label{sec:observations}

We analyze 12 hours of AIA observations of \AR{} NOAA 1158 in six EUV channels, 94, 131, 171, 193, 211, and 335 \AA{}, beginning at 2011 February 12 12:00:00 UTC and ending at 2011 February 13 00:00:00 UTC.
The \AR{} was chosen from the catalogue of \AR s originally compiled by \citet{warren_systematic_2012} and later studied by \citetalias{viall_survey_2017}.
The full-disk, level-1 AIA data products in FITS file format are obtained from the Joint Science Operations Center \citep[JSOC,][]{couvidat_observables_2016} at the full instrument cadence of 12 s and full spatial resolution using the drms Python client \citep{glogowski_drms_2019}.
This amounts to a total of 21597 images across all six channels and the entire 12 h observing window.

\begin{figure*}
    \centering
   \includegraphics[width=\columnwidth]{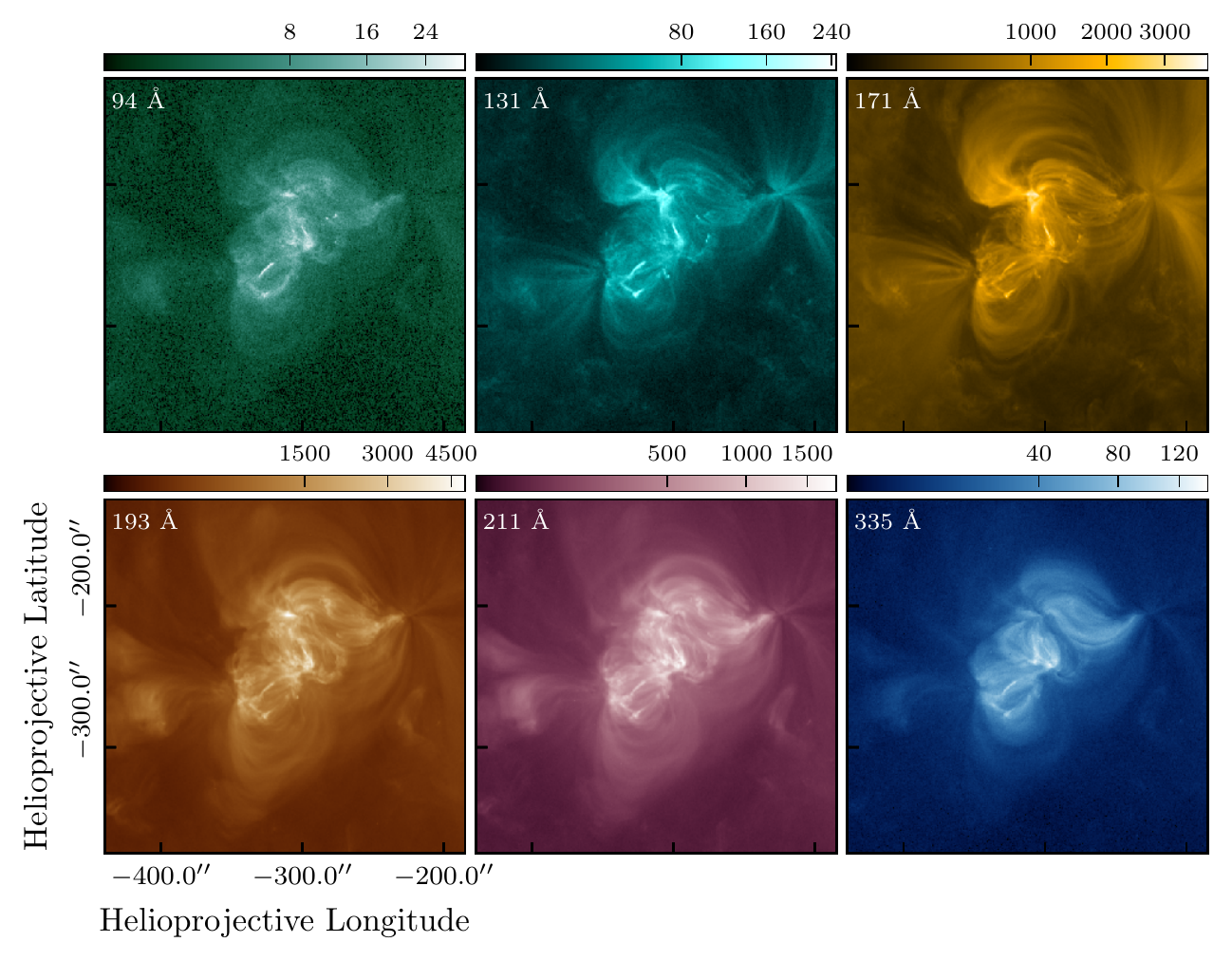}
    \caption{Active region NOAA 1158 as observed by AIA on 2011 February 12 15:32 UTC in the six EUV channels of interest. The data have been processed to level-1.5, aligned to the image at 2011 February 12 15:33:45 UTC, and cropped to the area surrounding NOAA 1158. The intensities are in units of DN pixel$^{-1}$ s$^{-1}$. In each image, the colorbar is on a square root scale and is normalized between zero and the maximum intensity. The color tables are the standard AIA color tables as implemented in sunpy.}
    \label{fig:intensity-maps}
\end{figure*}

After downloading the data, we apply the \texttt{aiaprep} method, as implemented in sunpy \citep{the_sunpy_community_sunpy_2020}, to each full-disk image in order to remove the instrument roll angle, align the center of the image with the center of the Sun, and scale each image to a common spatial resolution such that images in all channels have a spatial scale of 0.6\arcsec-per-pixel.
Additionally, we normalize each image by the exposure time such that the data have units of DN pixel$^{-1}$ s$^{-1}$.
Next, we align each image with the observation at 2011 February 12 15:33:45 UTC (the time of the original observation of NOAA 1158 by \citet{warren_systematic_2012}) by ``derotating'' each image using the Snodgrass empirical rotation rate \citep{snodgrass_magnetic_1983}.
After aligning the images in every channel to a common time, we crop each full-disk image such that the bottom left corner of the image is $(-440\arcsec,-375\arcsec)$ and the top right corner is $(-140\arcsec,-75\arcsec)$, where the two coordinates are the longitude and latitude, respectively, in the helioprojective coordinate system \citep[see][]{thompson_coordinate_2006} defined by an observer at the location of the SDO spacecraft on 2011 February 12 15:33:45.
\autoref{fig:intensity-maps} shows the level-1.5, exposure-time-normalized, derotated, and cropped AIA observations of \AR{} NOAA 1158 at 2011 February 12 15:33:45 in all six EUV channels of interest.


\subsection{Emission Measure Slopes and Peak Temperatures}\label{sec:em_slopes}


After prepping, aligning, and cropping all 12 h of AIA data for all six channels, we carry out the same analysis that we applied to our predicted observations in \citetalias{barnes_understanding_2019} in order to compute the diagnostics of the heating: the emission measure slope, peak temperature, time lag, and maximum cross-correlation.
First, we compute the emission measure distribution, \dem, in each pixel of the \AR{} from the time-averaged intensities from all six channels using the regularized inversion method of \citet{hannah_differential_2012}.
As in \citetalias{barnes_understanding_2019}, we use temperature bins of width $\Delta\log T=0.1$ with the left and right edges at $10^{5.5}$ K and $10^{7.2}$ K, respectively.
The uncertainties on the intensities are estimated using the \texttt{aia\_bp\_estimate\_error.pro} procedure provided by the AIA instrument team in the SolarSoftware package \citep[SSW,][]{freeland_data_1998}.  


\begin{figure*}
    \centering
   \includegraphics[width=\columnwidth]{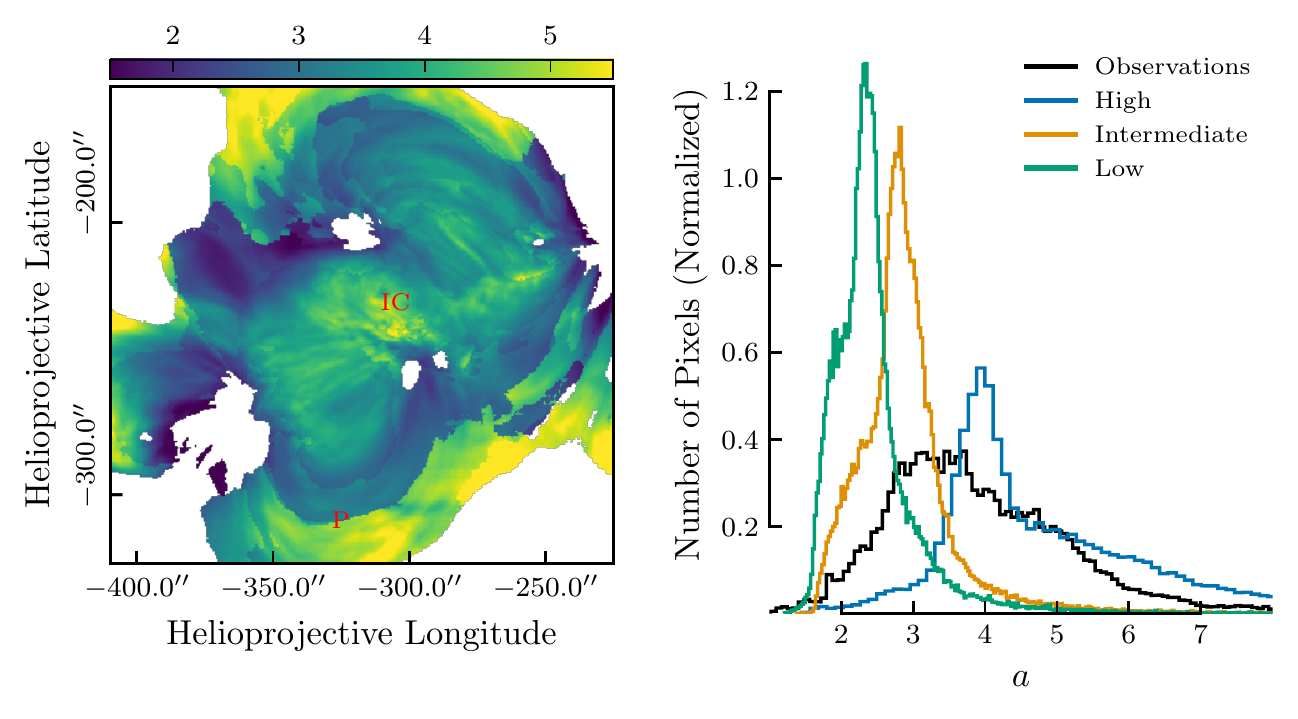}
    \caption{\textit{Left:} Map of emission measure slope, $a$, in each pixel of \AR{} NOAA 1158. The \dem{} is computed from the observed AIA intensities in the six EUV channels time-averaged over the 12 h observing window. The \dem{} in each pixel is then fit to $T^a$ over the temperature interval $8\times10^5\,\textup{K}\le T < T_{peak}$. Any pixels with $r^2<0.75$ are masked and colored white. ``IC'' and ``P'', marked in red, denote the inner core and periphery of the active region, respectively. \textit{Right:} Distribution of emission measure slopes from the left panel (black) and from \citetalias{barnes_understanding_2019} (blue, orange, green). Each histogram is normalized such that the area under the histogram is equal to 1.}
    \label{fig:em-slopes}
\end{figure*}

The left panel of \autoref{fig:em-slopes} shows the emission measure slope, $a$, as computed from the observed emission measure distribution in each pixel of \AR{} NOAA 1158.
We calculate $a$ by fitting a first-order polynomial to the log-transformed emission measure and the temperature bin centers, $\log_{10}\textup{EM}\sim a\log_{10}T$.
As in \citetalias{barnes_understanding_2019}, the fit is only computed over the temperature range $8\times10^5\,\textup{K}\le T \le T_{peak}$, where $T_\textup{peak}=\argmax_T\,\textup{EM}(T)$ is the temperature at which the emission measure distribution peaks.
If $r^2<0.75$ in any pixel, where $r^2$ is the correlation coefficient for the first-order polynomial fit, the pixel is masked and colored white. 

Similar to \citet{barnes_understanding_2019}, we define the \textit{inner core} as the area near the center of the \AR{} whose X-ray and EUV emission is dominated by short, closed loops.
We define the \textit{periphery} as the region farthest from the inner core containing closed loops which are still visible relative to the surrounding diffuse emission.
The inner core (``IC'') and the periphery (``P'') of the \AR{} are indicated in red in the left panel of \autoref{fig:em-slopes}.
While these definitions of different parts of the \AR{} are conceptual and qualitative rather than strictly quantitative, they will be useful in discussing our results in \autoref{sec:compare} and \autoref{sec:discussion}.

The emission measure slope tends to be more steep near the center of the \AR{} and tends to increase from $\sim2.5$ to $>5$ moving from the periphery to the inner core of the \AR{}.
This result is consistent with \citet{del_zanna_evolution_2015} who computed the emission measure slope in each pixel of \AR{} NOAA 1193 and found that $a$ was greatest near the middle of the \AR{}.
The exception to this trend is the spatially-coherent structure on the lower edge of the \AR{}, near $(-300\arcsec,-320\arcsec)$, which shows emission measure slopes $>5$.
A few regions on the top edge, near $(-350\arcsec,-140\arcsec)$, also show higher emission measure slopes.

The right panel of \autoref{fig:em-slopes} shows the distribution of emission measure slopes for every pixel in the \AR{} where $r^2\ge0.75$.
As noted in the legend, the black histogram denotes the observed slopes while the blue, orange, and green histograms are the distributions of emission measure slopes computed from the predicted AIA intensities in \citetalias{barnes_understanding_2019} for high-, intermediate-, and low-frequency nanoflares, respectively.
The mean of the observed distribution of $a$ is 3.92
and the standard deviation is 1.30. 

We find that the observed distribution of slopes overlaps the distributions of predicted slopes for all three heating scenarios, suggesting that no single heating scenario can explain the width of the distribution and that a range of nanoflare heating frequencies is operating across the \AR.
In particular, the observed distribution of slopes overlaps quite strongly with both the intermediate- and high frequency-slopes.
Compared to the simulated distributions of $a$ for low- and intermediate-frequency heating, the observed distribution is wide with a relatively flat top between $a\approx3$ and $a\approx4$.
In contrast to all three simulated distributions, the observed slope distribution is not strongly peaked about any single value of $a$.


\begin{figure}
    \centering
   \includegraphics[width=0.5\columnwidth]{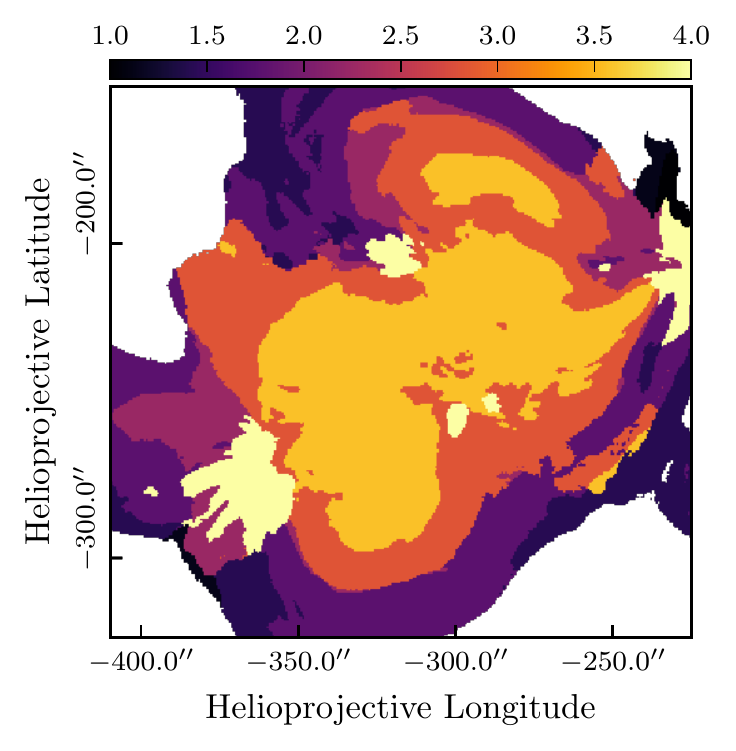}
    \caption{Map of $T_{peak}$, the center of the temperature bin at which the emission measure distribution peaks, in MK, in each pixel of \AR{} NOAA 1158. Any pixels with total emission measure $<10^{27}$ cm$^{-5}$ are masked and colored white. The colorbar is linearly spaced between 1 and 4 MK.}
    \label{fig:em-tpeaks}
\end{figure}

Additionally, we measure $T_{peak}$, the value at the center of the temperature bin in which the emission measure distribution is maximized, in each pixel of \AR{} NOAA 1158 from the derived emission measure distributions.
\autoref{fig:em-tpeaks} shows $T_{peak}$, in MK, in each pixel of the \AR{}.
We find that the the majority of pixels have values of $T_{peak}$ between 1.5 and 3.5 MK.
Additionally, we find that $T_{peak}$ is highest in the inner core of the \AR{}, just over 3.5 MK, and decreases to between 1.5 and 2 MK as we move outward toward the periphery.
Though not shown here, we find this same general trend in our model \AR s for all heating frequencies.
For example, Figure 5 of \citetalias{barnes_understanding_2019} shows that the \dem{} distributions near the inner core of our model \AR{} have $T_{peak}\approx3$ MK while the \dem{} closer to the periphery has $T_{peak}\approx2$ MK.

We note that there are several small regions which have $T_{peak}\ge4$ MK.
Comparing \autoref{fig:em-tpeaks} with \autoref{fig:intensity-maps}, we see that these regions correspond to ``open'' fan loops which are cooler and have lower signal-to-noise ratio, suggesting that the inverted \dem{} solutions are not reliable in these regions.
However, we already exclude the emission measure slopes in these regions due to their low correlation coefficients (see \autoref{fig:em-slopes}) and thus the unreliable inverted solutions will not affect our later predictions in \autoref{sec:compare}.

\subsection{Time Lags}\label{sec:timelags}


\begin{figure}
    \centering
   \includegraphics[width=0.75\columnwidth]{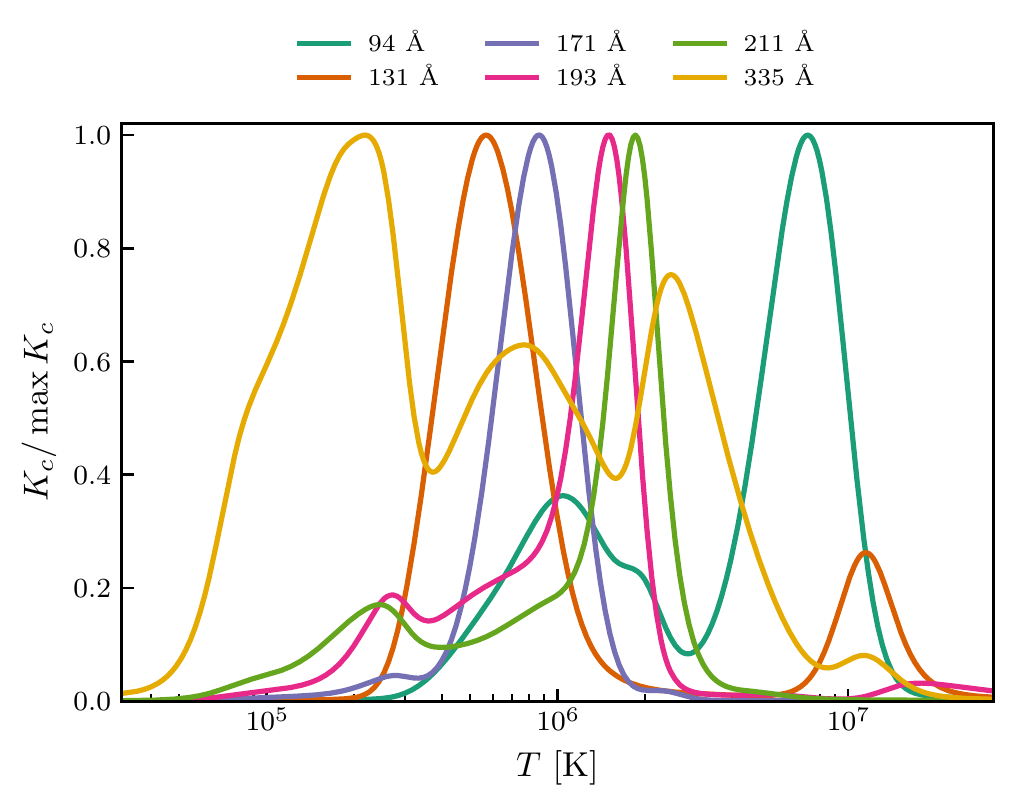}
    \caption{Temperature response functions for all six EUV channels of AIA as a function of temperature computed by the \texttt{aia\_get\_response.pro} procedure in SSW. Each response function is normalized to its maximum.}
    \label{fig:aia-response}
\end{figure}

Next, we apply the time lag analysis of \citet{viall_evidence_2012} to every pixel in the \AR{} over the entire 12 h observing window at the full temporal and spatial resolution.
As in \citetalias{barnes_understanding_2019}, we compute the cross-correlation, $\mathcal{C}_{AB}$, between all possible ``hot-cool'' pairs, $AB$, of the six EUV channels of AIA (15 in total) and find the time lag, $\tau_{AB}$, the temporal offset which maximizes the cross-correlation, in each pixel of the observed \AR{}.
We consider all possible offsets over the interval $\pm6$ h.
Following the convention of \citet{viall_evidence_2012}, we take the order of the channels, from hottest to coolest, to be: 94, 335, 211, 193, 171, 131 \AA{}, meaning that \textit{a positive time lag indicates cooling plasma}.
Observationally, this is often a good representation of how the plasma in a quiescent \AR{} evolves through the AIA channels.
The response curves as a function of temperature for these six EUV channels of AIA are shown in \autoref{fig:aia-response}.
The details of the cross-correlation and time lag calculations can be found in the appendix of \citetalias{barnes_understanding_2019}.


\begin{figure*}
    \centering
   \includegraphics[width=\columnwidth]{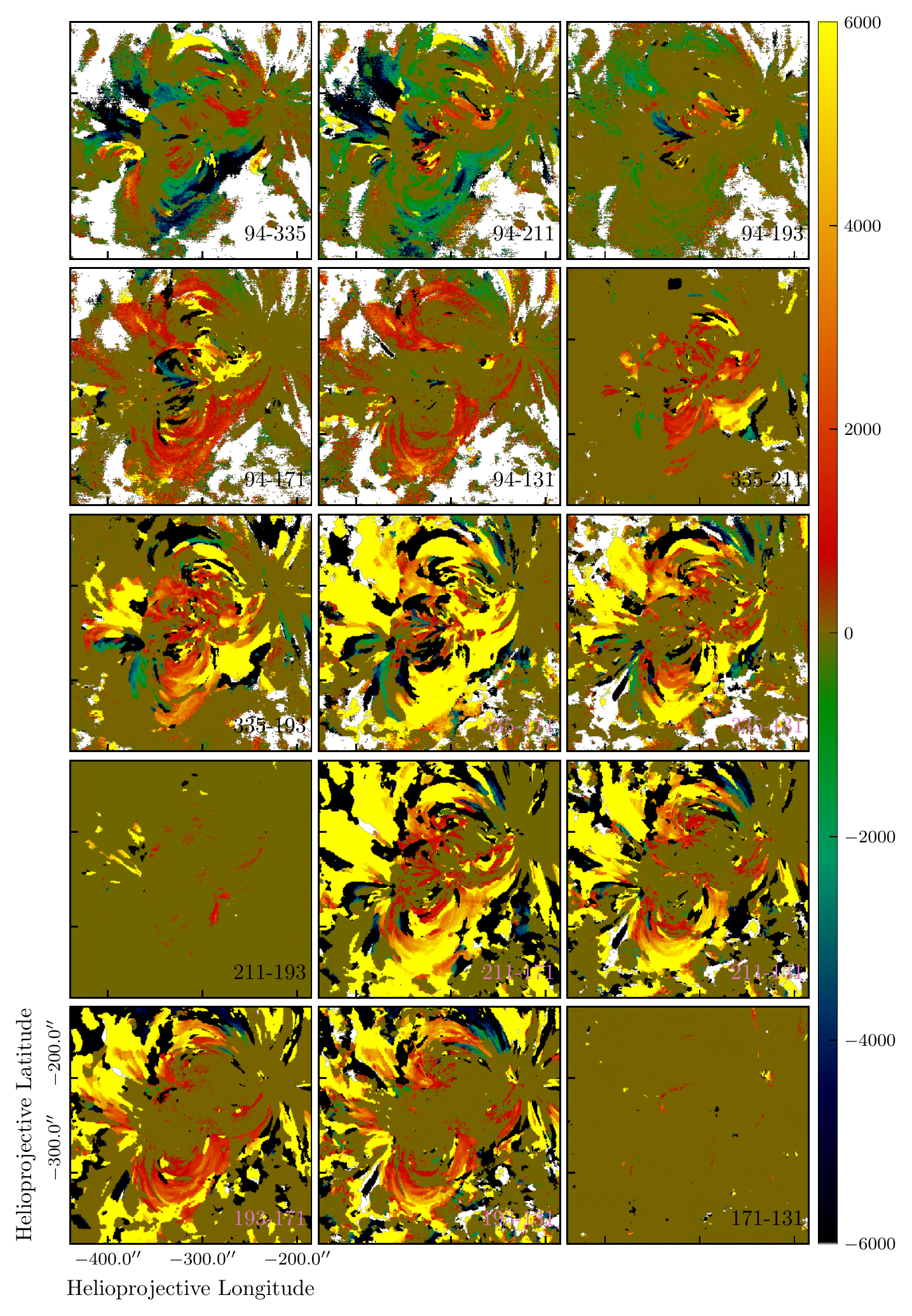}
    \caption{Time lag maps of \AR{} NOAA 1158 for all 15 channel pairs. The value of each pixel indicates the temporal offset, in seconds, which maximizes the cross-correlation \citepalias[see Appendix C of][]{barnes_understanding_2019}. The range of the colorbar is $\pm6000$ s. If $\max\mathcal{C}_{AB}<0.1$, the pixel is masked and colored white. Each map has been cropped to emphasize the core of the \AR{} such that the bottom left corner and top right corner of each image correspond to $(-440\arcsec,-380\arcsec)$ and $(-185\arcsec,-125\arcsec)$, respectively.}
    \label{fig:timelags}
\end{figure*}

\autoref{fig:timelags} shows the time-lag maps of \AR{} NOAA 1158 for all 15 channel pairs.
Blacks, blues, and greens indicate negative time lags while reds, oranges, and yellows correspond to positive time lags.
Olive green denotes near zero time lag.
The range of the colorbar is $\pm6000$ s.
If the maximum cross-correlation in a given pixel is too small, $\max\mathcal{C}_{AB}<0.1$, the pixel is masked and colored white.

Note that \citetalias{viall_survey_2017} carried out the time lag analysis on this same \AR{}, NOAA 1158 (their region 2), as part of a survey of the catalogue of \AR{}s compiled by \citet{warren_systematic_2012}.
We repeat this analysis here to ensure that we are treating the observed intensities in the exact same manner as the predicted intensities from \citetalias{barnes_understanding_2019}.
The method employed here for calculating the time lag \citepalias[see Appendix C of][]{barnes_understanding_2019} yields quantitatively identical results to that used by \citetalias{viall_survey_2017} \citep[see Section 2 of][]{viall_evidence_2012}.
Comparing \autoref{fig:timelags} to Figure 2 and Figure 4 of \citetalias{viall_survey_2017}, we find all of the same qualitative features in the time lag maps for each channel pair.
We note that, while there are differences between the two sets of time lag maps, these can be attributed to differences in the field of view of the cutouts and ``derotation'' reference date.
Additionally, \citetalias{viall_survey_2017} did not apply any masking based on the maximum cross-correlation value.

For the majority of the channel pairs, we find persistent positive time lags across most of the \AR{}, indicative of plasma cooling through the AIA passbands.
The 94-131, 211-131, 193-171, and 193-131 \AA{} pairs show coherent positive time lags on the periphery of the \AR{}, but zero time lag in the center of the \AR{}.
On the other hand, 171-131 \AA{} channel pair map shows zero time lags in nearly every pixel of the \AR{}.
While the 211-193 \AA{} channel pair map also appears to show mostly zero time lags, there are a significant number of positive time lags compared to the 171-131 \AA{} channel pair, consistent with \citet{viall_evidence_2012,viall_survey_2017}.
From \autoref{fig:aia-response}, we see that both of these channel pairs are strongly overlapping in temperature space such that their respective peaks in intensity are likely to be close to coincident in time as the plasma cools.
However, the presence of the positive, though small 211-193 \AA{} time lags compared to the zero 171-131 \AA{} time lags is indicative of plasma cooling into, but not through the 131 \AA{} channel \citep{bradshaw_patterns_2016}.

Additionally, the 94-335, 94-193, and 94-211 \AA{} pairs all show significant coherent negative time lags.
Because the 94 \AA{} channel is bimodal in temperature (see \autoref{fig:aia-response}), a negative time lag is indicative of the plasma cooling first through the ``cool'' channel and then through the cooler component of the 94 \AA{} bandpass.
In this case, the cooler, 1 MK component of the 94 \AA{} dominates the emission.
The 94-171 and 94-131 pairs show only positive time lags because the 171 \AA{} and 131 \AA{} channels peak at cooler temperatures than the cool component of the 94 \AA{} channel.
See \citetalias{viall_survey_2017} for a more detailed discussion of the time lag results from NOAA 1158.
Note that unlike the predicted time lags in \citetalias{barnes_understanding_2019}, none of the pairs involving the 131 \AA{} channel, which is also bimodal in temperature, show any coherent negative time lags and, in particular, the inner cores of each 131 \AA{} pair show zero time lag.
This is indicative of an excess of hot plasma in our model \AR{} relative to the observations.


\begin{figure*}
    \centering
   \includegraphics[width=\columnwidth]{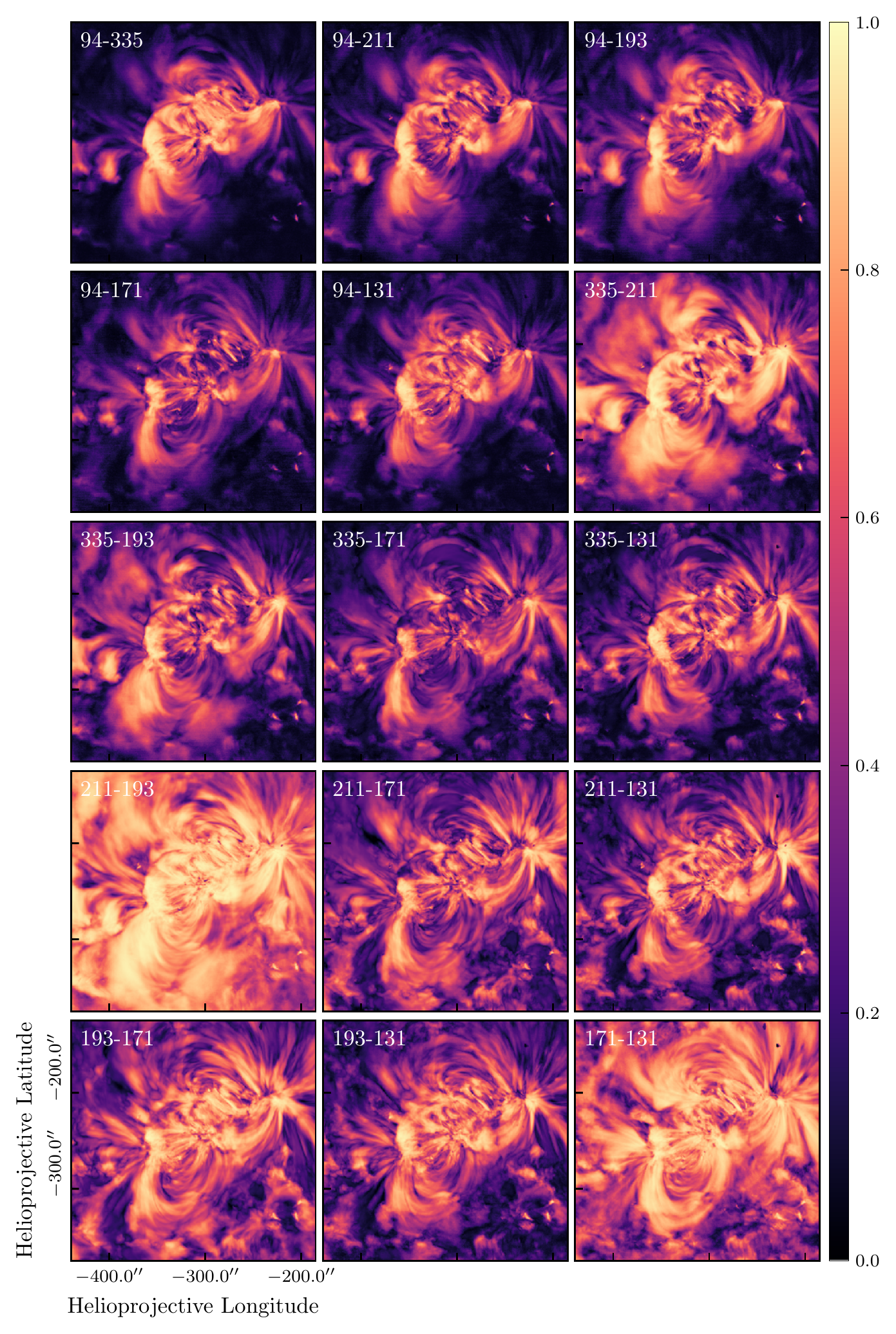}
    \caption{Same as \autoref{fig:timelags} except here we show the maximum value of the cross-correlation, $\max\mathcal{C}_{AB}$, for each channel pair.}
    \label{fig:correlations}
\end{figure*}

\autoref{fig:correlations} shows the maximum cross-correlation, $\max\mathcal{C}_{AB}$, in each pixel of the \AR{}.
In this figure, we do not mask any of the pixels. Though the value of the cross-correlation can range from $-1$ (perfectly anti-correlated) to $+1$ (perfectly correlated), the colorbar only ranges from 0 to 1 as we are only interested in whether the light curves in each channel pair are in phase.
In practice, these values are rarely less than zero as the time lag method finds the maximum cross-correlation over all possible time lags.

In every channel pair, we find that the maximum cross-correlation maps reveal coherent loop-like structures similar to those seen in the observed intensities shown in \autoref{fig:intensity-maps}, indicating that these loops and the surrounding diffuse emission are evolving coherently through the AIA passbands.
In most channel pairs, the inner core tends to have the highest cross-correlation while areas near the corners of the images have low cross-correlation.
In general, this is expected as the periphery has less emission than the core in all channels and thus lower count rates.
This lower signal-to-noise ratio means that any physical variations are more likely to be lost in the noise.
Note that the channel pairs which had the most zero time lags in \autoref{fig:timelags}, 211-193 and 171-131, show high cross-correlations across the entire \AR{}.
This again emphasizes the point that zero time lags do not correspond to steady heating.
If the whole \AR{} was producing steady emission, we would expect low cross-correlation and no preferred time lag due to the photon noise dominating the variability in each channel \citep{viall_signatures_2016}.
 %
\section{Classification Model}\label{sec:compare}


Rather than manually comparing our observations and simulations using all of the aforementioned diagnostics, we systematically assess our observations of NOAA 1158 in terms of the heating frequency by training a random forest classifier comprised of many decision tress on our predicted observables from \citetalias{barnes_understanding_2019}.
We then use our trained model to classify each observed pixel in terms of high-, intermediate-, or low-frequency heating as defined in \autoref{eq:heating_types}.
Unlike more traditional statistical methods, this approach allows us to simultaneously consider an arbitrarily large number of features when deciding which frequency best fits the observation.
In the parlance of statistical learning, the heating frequency (low, intermediate, or high) is the \textit{class}, the emission measure slope, peak temperature, time lags, and cross-correlations are the \textit{features}, and the pixels are the \textit{samples}.

Following the explanation of \citet[chapter 8]{james_introduction_2013}, a decision tree recursively partitions the feature space of interest into a set of terminal nodes, or leaves, using a top-down, ``greedy'' approach called recursive binary splitting.
At each node in the tree, a feature and an associated split point are chosen to maximize the number of observations of a single class in the resulting nodes.
A common measure of the homogeneity or \textit{purity} of each node is the Gini index,
\begin{equation}\label{eq:gini-index}
    G_m = \sum_k \hat{p}_{mk} (1 - \hat{p}_{mk}),
\end{equation}
where $k$ indexes the class, $m$ indexes the node, and $\hat{p}_{mk}$ is the proportion of the observations at node $m$ that belong to class $k$.
Note that as the purity of $m$ increases (i.e. $\hat{p}_{mk}\to0,1$), $G_m$ decreases ($G_m\to0$).
Alternative measures of node purity may also be used \citep[see section 9.2.3 of][]{hastie_elements_2009}.
For every resulting terminal node in the tree, the assigned class is determined by the most commonly occurring class of every observation at that node.
In the case of our application presented here, a node will be assigned the label ``high'', ``intermediate'', or ``low'' frequency depending on the majority heating frequency label of the model pixels used for training that end on that node.

Decision trees are commonly used in classification problems because they are computationally efficient and relatively easy to interpret.
Unlike many statistical learning techniques, decision trees do not assume any functional mapping between the inputs and outputs such that arbitrary, non-linear relationships can be learned by the model.
However, decision trees have two primary weaknesses: (1) they are known to have lower predictive accuracy than other more restrictive classification strategies and (2) they have high variance such that a single tree is not very robust to small changes in the training data \citep{james_introduction_2013}.

While individual decision trees are ``weak learners,'' combined they give accurate and robust predictions.
Random forest classifiers, first developed by \added{\citet{ho_random_1995} and later improved by} \citet{breiman_random_2001}, provide an ensemble statistical learning method for combining many noisy, decorrelated decision trees in order to improve prediction accuracy and robustness.
As in the bootstrap-aggregation, or ``bagging'', technique developed by \citet{breiman_bagging_1996}, each tree in the random forest is trained on only a subset of the total training data in order to reduce the variance of the model.
Additionally, at each node in each tree, a random subset of the total features are considered as candidates for splitting in order to decrease the correlation between trees.
A typical rule-of-thumb is to consider only $\left\lfloor\sqrt{p}\right\rfloor$ features at each split, where $p$ is the total number of features.
This further reduces the variance and prevents a single feature from dominating the decision in every tree.
Once each tree in the forest has been built using the training data, an unlabeled observation is classified by traversing each tree in the forest and taking the majority vote of the class at the terminal node of each tree.
See \citet[chapter 15]{hastie_elements_2009} for a detailed discussion of random forests for both classification and regression.

We note that random forests have recently been successfully applied in other areas of solar physics, namely predicting the properties and occurrence of flares.
For example, \citet{campi_feature_2019} employed two prediction methods, including random forests, to predict the occurrence of flares using the magnetic properties of the candidate \AR s.
In particular, they exploited the ability of a random forest to provide information about which properties were most important in classifying an \AR{} as flaring or non-flaring.
This concept, called ``feature importance,'' is discussed in more detail in \autoref{sec:feature-importance}.
Additionally, \citet{reep_forecasting_2021} recently used a random forest regressor to predict the remaining duration of a flare.
They found that, given a set of parameters derived from the peak X-ray flux as observed by the GOES satellite, they can predict the remaining duration of the flare with significantly greater accuracy than traditional methods that rely on simple linear regression.

\subsection{Data Preparation and Model Parameters}\label{sec:data-prep}

To build our classification model, we use the random forest classifier as implemented in the scikit-learn package for machine learning in Python \citep{pedregosa_scikit-learn_2011}.
Using the predicted peak temperatures, emission measure slopes, time lags, and maximum cross-correlations from \citetalias{barnes_understanding_2019}, we train a single random forest classifier composed of 500
trees each with a maximum depth of 30.
At each node, $\left\lfloor\sqrt{32}\right\rfloor=5$ possible split candidates are randomly selected from the $15\,\textup{time lags} + 15\,\textup{cross-correlations} + 1\,\textup{peak temperature} + 1\,\textup{emission measure slope}=32$ total features.
\added{While \citetalias{barnes_understanding_2019} only discussed 31 synthetic features (15 time lags, 15 maximum cross-correlation values, 1 emission measure slope), here we have calculated $T_{peak}$ for our synthetic emission measure distributions as well such that we have 32 features in our synthetic data set and 32 in our observational data set.}
We note that all of these features are likely to be correlated with one another to some extent.
In \autoref{sec:feature-combos}, we examine how training our model on different combinations of these features impacts our predictions.

Before training the model, we flatten the predicted emission measure slope, peak temperature, time lag, and cross-correlation maps from \citetalias{barnes_understanding_2019} for the high-, intermediate-, and low-frequency heating cases into an array of length $n_xn_y$, where $n_x$ and $n_y$ are the dimensions of the predicted images.
We stack each flattened array column-wise in features and row-wise in heating frequency such that all of the simulated data are encapsulated in a single data matrix $X$ of dimension $n\times p$.
$p=32$ is the total number of features and $n=3n_xn_y - n_\textup{mask}=110619$ is the total number of pixels for all heating frequencies minus those pixels which were masked in at least one feature of one frequency.

As before, we mask pixels where $r^2<0.75$ for the emission measure slope fit and where $\max\mathcal{C}_{AB}<0.1$ for the cross-correlation.
If a pixel is masked in any frequency case, we mask it in all other frequencies to ensure that we have an equal number of high-, intermediate-, and low-frequency data points.
This ensures we do not have an imbalance in the number of points in each class when training our model.
In total, we mask 85.25$\%$ of the total pixels from all three heating frequency cases for our simulated features used for training our classification models.
The heating frequency label or class is numerically encoded as 0 (high), 1 (intermediate), or 2 (low) and similarly stacked to create a single response vector $Y$ of dimension $n\times1$.

We apply a $2/3-1/3$ train-test split to $X$ and $Y$ such that approximately $1/3$ of the samples are reserved for model evaluation to ensure that our model has not overfit the data\footnote{As an additional check, we applied random permutations cross-validation for five different iterations and found that misclassification error on each resulting test set was comparable to the out-of-bag error computed by the random forest classifier for a single $2/3-1/3$ split.}.
This produces four separate matrices: $X_\textup{train},Y_\textup{train},X_\textup{test},Y_\textup{test}$.
The data are not centered to a mean of 0 or scaled to unit standard deviation.
By transforming the data in this manner, we are treating each pixel in the image as an independent sample with $p$ associated features per sample.

The same procedure as described above is applied to the observed emission measure slopes, peak temperatures, time lags, and cross-correlations as shown in \autoref{fig:em-slopes}, \autoref{fig:em-tpeaks}, \autoref{fig:timelags}, and \autoref{fig:correlations}, respectively.
These results are flattened to a single data matrix $X^\prime$ of dimension $n^\prime\times p$, where $n^\prime=n_x^\prime n_y^\prime - n^\prime_\textup{mask}=69796$.
We apply the same masking procedure as described above based on the measured emission measure slopes and maximum cross-correlations and mask 72.08$\%$ of the total observed pixels.
The random forest model is trained on $X_\textup{train},Y_\textup{train}$ and model performance is evaluated on the ``unseen'' test set $X_\textup{test},Y_\textup{test}$.
The trained model is then applied to $X^\prime$ in order to predict the heating frequency in each pixel, $Y^\prime$.

Though model ``hyperparameters'' (e.g. the number of estimators, maximum tree depth, number of split candidates) are often determined systematically by finding the set of parameters that minimize the misclassification error on a subset of the test data, we do not apply these more formal procedures here.
A manual exploration of the hyperparameters reveals that adding more than 500
trees to the random forest provides only a marginal decrease in the test error while increasing the training time.
Similarly, we find a maximum depth of 30
for each decision tree provides sufficient complexity to each tree as evaluated by the test error while not significantly increasing the computational cost of the training.
However, in case A (see \autoref{tab:cases}), we find that less complex trees (i.e. lower maximum depth) result in a reduction in the misclassification error by $7-8\%$.

\subsection{Different Feature Combinations}\label{sec:feature-combos}

\begin{deluxetable*}{ccccccc}
\tablecaption{The four different combinations of emission measure slope, peak temperature, time lag, and maximum cross-correlation. The third column lists the total number of features used in the classification. The fourth column gives the misclassification error as evaluated on $X_\textup{test},Y_\textup{test}$. The fifth, sixth, and seventh columns show the percentage of pixels labeled as high-, intermediate-, and low-frequency heating, respectively.\label{tab:cases}}
\tablehead{\colhead{Case} & \colhead{Parameters} & \colhead{$p$} & \colhead{Error} & \colhead{High} & \colhead{Inter.} & \colhead{Low}}
\startdata
A & $a,T_{peak}$ & 2 & 0.25 & 0.473 & 0.354 & 0.173 \\
B & $\tau_{AB},\mathcal{C}_{AB}$ & 30 & 0.03 & 0.832 & 0.115 & 0.053 \\
C & $a,T_{peak},\tau_{AB},\mathcal{C}_{AB}$ & 32 & 0.02 & 0.736 & 0.235 & 0.029 \\
D & Top 10 features from \autoref{tab:importance} & 10 & 0.05 & 0.671 & 0.274 & 0.055
\enddata
\end{deluxetable*}

We apply the train-test-predict procedure described above to all four cases listed in \autoref{tab:cases}.
In case A, the random forest classifier is trained only on the features related to the emission measure distribution, the emission measure slope, $a$, and the peak temperature, $T_{peak}$, such that the $X$ and $X^\prime$ have dimensions $n\times2$ and $n^\prime\times2$, respectively.
In case B, the classifier is trained on the time lags and maximum cross-correlations for all 15 channel pairs for a total of $p=30$ features while in case C, every feature (emission measure slope, peak temperature, 15 time lags, 15 maximum cross-correlations) is used such that $p=32$.
We discuss case D in \autoref{sec:feature-importance}.
The fourth column in \autoref{tab:cases} lists the misclassification error as evaluated on the test set, $X_\textup{test},Y_\textup{test}$ and the fifth, sixth, and seventh columns show the fraction of pixels classified as high-, intermediate-, and low-frequency, respectively.


\begin{figure*}
    \centering
   \includegraphics[width=\columnwidth]{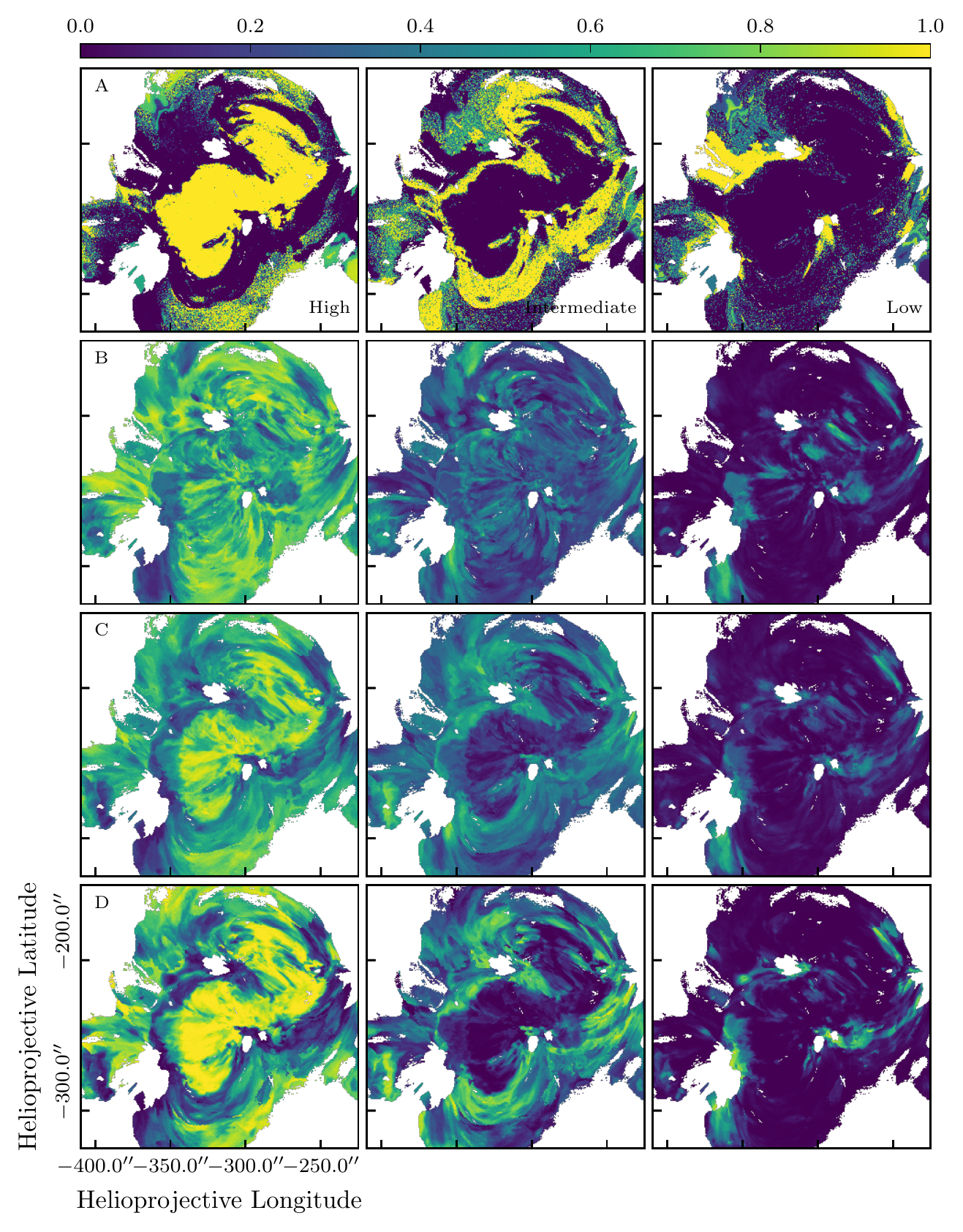}
    \caption{Classification probability for each pixel in the observed \AR{}. The rows denote the different cases in \autoref{tab:cases} and the columns correspond to the different heating frequency classes. If any of the 32 features is masked in a particular pixel, the pixel is colored white. Note that summing over all heating probabilities in each row gives 1 in every pixel.}
    \label{fig:probability-maps}
\end{figure*}

After computing the predicted heating frequency for $X^\prime$, the resulting classifications, $Y^\prime$, are mapped back to the corresponding observed pixel locations to create a map of the heating frequency.
\autoref{fig:probability-maps} shows the probability that each pixel corresponds to a particular heating frequency.
The rows denote the different feature subsets as given in \autoref{tab:cases} and the columns correspond to the different heating frequency classes.
The class probability, as computed by the scikit-learn package, in each pixel is the mean class probability of all trees in the random forest classifier.
The class probability for an individual tree is the proportion of all training samples at the terminal node that belong to that class.
\added{The class probability in each pixel provides a measure of the confidence of the classifier in assigning a given heating frequency label to that pixel.
If one heating frequency has a class probability close to 1, the other two classes will be, by definition, close to 0, and the confidence of the classification is high.
If the class probability for each heating frequency is $1/3$, the classifier is not able to distinguish which heating frequency is most probable in that pixel.}


\begin{figure*}
    \centering
   \includegraphics[width=\columnwidth]{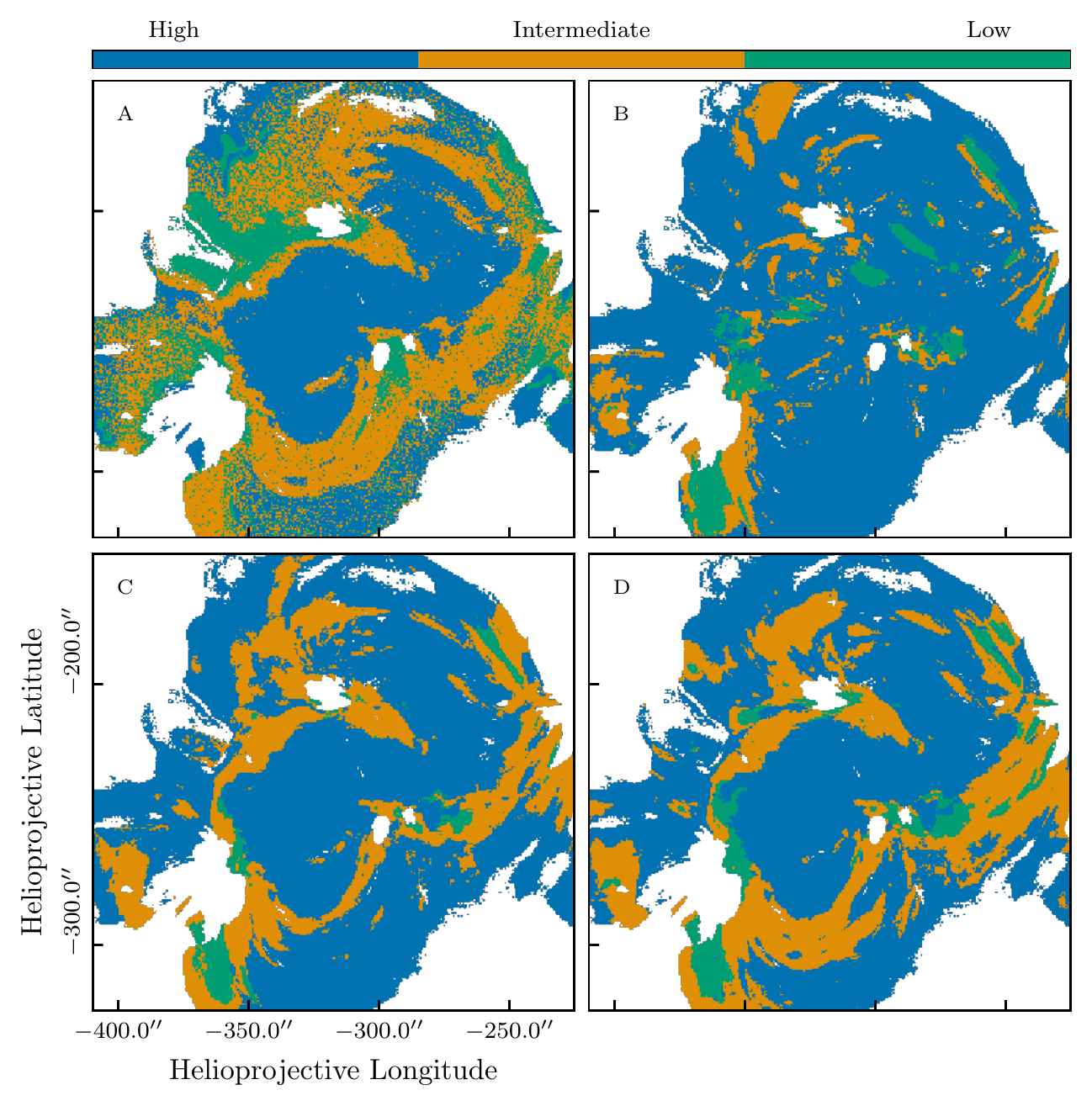}
    \caption{Predicted heating frequency classification in each pixel of NOAA 1158 for each of the cases in \autoref{tab:cases}. The classification is determined by which heating frequency class has the highest mean probability over all trees in the random forest. Each pixel is colored blue, orange, or green depending on whether the most likely heating frequency is high, intermediate, or low, respectively. If any of the 32 features is masked in a particular pixel, the pixel is colored white.}
    \label{fig:frequency-maps}
\end{figure*}

\autoref{fig:frequency-maps} shows the heating frequency, or class, as predicted by the random forest classifier in each pixel of the observed \AR{} for all four cases in \autoref{tab:cases}.
The predicted class is the one which has the highest mean probability as computed over all trees in the random forest.
Each pixel is colored blue, orange, or green depending on whether the class with the highest mean probability is high-, intermediate-, or low-frequency, respectively.

We find that for each combination of features in \autoref{tab:cases}, high-frequency heating dominates at the center of the \AR{}.
This result is consistent with \AR{} core observations of hot, steadier emission \citep{warren_evidence_2010,warren_constraints_2011}, steep emission measure slopes \citep[e.g.][]{winebarger_using_2011,del_zanna_evolution_2015}, and lack of variability in the velocity \citep{brooks_flows_2009} near the loop footpoints.
The frequency classification map for case A is as expected given the observed emission measure slope map in the left panel of \autoref{fig:em-slopes} and the well-separated distributions of emission measure slopes from the different heating frequencies as shown in the right panel of \autoref{fig:em-slopes}.

From the fourth column of \autoref{tab:cases}, we find that adding more features to the classifier significantly improves the accuracy as computed on the test data set.
However, comparing frequency maps in cases A ($p=2$) and C ($p=32$), we find that the general pattern of heating frequency across the \AR{} is similar despite the large differences between the misclassification error in case A (0.25) and case C (0.02).
Additionally, looking at the seventh column of \autoref{tab:cases} and the panels in the third column of \autoref{fig:probability-maps}, we find that adding the time lag and maximum cross-correlation features significantly decreases the number of pixels classified as low frequency.
Overall, we find that training the classifier on all of the features (C) versus only the emission measure slope and peak temperature (A) increases the number of pixels classified as high-frequency and decreases the number of intermediate- and low-frequency pixels, with the reduction in the low-frequency pixels being the most prominent.

Interestingly, we find that when we use relatively shallow trees to build the random forest (e.g. a maximum depth of $<10$), the misclassification error on the test data set in case B becomes larger than that of case A, despite $p_\textup{B}>p_\textup{A}$.
If very complex trees (maximum depth $>100$) are used in case A, the model overfits the data and the resulting classification becomes very noisy. 
However, in case B (and C), increasing the maximum depth continually decreases the test error, suggesting that the model is not overfitting the data. 
This seems to suggest that the relationship between the heating frequency and the time lag, as well as the maximum cross-correlation, is much more complex than that of the relationship between the heating frequency and the emission measure slope and the peak temperature.

\subsection{Feature Importance}\label{sec:feature-importance}

In addition to the predicted heating frequency, $Y^\prime$, for a set of features, $X^\prime$, it is also useful to know which of the $p$ features is most important in deciding to which class each observation (pixel) belongs.
One measure of the importance of each feature is the decrease in the Gini index,
\begin{equation}\label{eq:gini_gain}
    \Delta G_m = \frac{M_m}{M}\left( G_m - \frac{M_{m,R}}{M_m}G_{m,R} - \frac{M_{m,L}}{M_m}G_{m,L} \right),
\end{equation}
where $G_m$ is the Gini index as given by \autoref{eq:gini-index}, $M$ is the total number of samples in the tree, $M_m$ is the total number of samples at parent node $m$, $M_{m,R(L)}$ is the total number of samples at the right (left) child node, and $G_{m,R(L)}$ is the Gini index at the right (left) child node \citep{sandri_bias_2008}.
The importance of a particular feature in the random forest classification is then determined by summing \autoref{eq:gini_gain} over all nodes which split on that feature for every tree and averaging over all trees \citep{breiman_classification_1984}.

Note that if $G_{m,R}=G_{m,L}=G_m$, $\Delta G_m=0$ because the split at node $m$ did not improve the discrimination between classes compared to the split at the previous node.
However, if the purity of the left or right node increases such that $G_{m,R}$ or $G_{m,L}$ decrease relative to $G_m$, $\Delta G_m > 0$ because the split at node $m$ has added information to the classifier by preferentially sorting samples of a single class to either the left or right child node.
The larger the value of $\Delta G_m$, when summed over all relevant splits and averaged over all trees, the more important that particular feature is.

\begin{deluxetable}{ccc}
\tablewidth{\columnwidth}
\tablecaption{Ten most important features as determined by the random forest classifier in case C. The second column shows the variable importance as computed by \autoref{eq:gini_gain} and the third column, $\sigma$, is the standard deviation of the feature importance over all trees in the random forest. The second column is normalized such that the most important feature is equal to 1.\label{tab:importance}}
\tablehead{\colhead{Feature} & \colhead{Importance} & \colhead{$\sigma$}}
\startdata
$a$ & 1.0000 & 0.0776 \\
$\mathcal{C}_{211,193}$ & 0.4075 & 0.0892 \\
$\mathcal{C}_{193,171}$ & 0.3455 & 0.0838 \\
$\mathcal{C}_{211,171}$ & 0.2982 & 0.0854 \\
$T_{peak}$ & 0.1252 & 0.0140 \\
$\tau_{211,193}$ & 0.1043 & 0.0114 \\
$\mathcal{C}_{335,171}$ & 0.0948 & 0.0327 \\
$\tau_{171,131}$ & 0.0933 & 0.0141 \\
$\mathcal{C}_{335,211}$ & 0.0729 & 0.0185 \\
$\mathcal{C}_{335,193}$ & 0.0720 & 0.0248
\enddata
\end{deluxetable}

\autoref{tab:importance} shows the ten most important features from case C as determined by \autoref{eq:gini_gain} summed over all nodes in each tree and averaged over all trees.
The importance in the second column is normalized such that the most important feature is equal to 1.
In case D as listed in the last row of \autoref{tab:cases}, only these ten features are used to train the random forest classifier and classify each observed pixel.
The probability of each heating frequency for case D is shown in the last row of \autoref{fig:probability-maps} and the map of the most likely heating frequency in each pixel is shown in the bottom-right panel of \autoref{fig:frequency-maps}.
We find that the probability maps in the last row of \autoref{fig:probability-maps} and the frequency map in the bottom right panel of \autoref{fig:frequency-maps} reveal approximately the same patterns of heating frequency across the \AR{} as the maps for case C in which all 32 features were included.
Additionally, using less than $1/3$ of the total number of features, we achieve a misclassification error of 0.05, comparable to, though higher than, case C.

According to \autoref{tab:importance}, the emission measure slope, $a$, has the most discriminating power in the random forest classifier.
In particular, $a$ is more important than the second most important feature by over a factor of 2 and more important than the most important time lag feature by nearly an order of magnitude.
This is not surprising given that the value of the time lag and maximum cross-correlation measurements is in the combination of all channel pairs, allowing one to track the evolution of the plasma as it peaks in subsequently cooler channels.
This is confirmed by the lower test error in cases B and C compared to case A, as seen in \autoref{tab:cases}.
Additionally, we find that $T_{peak}$ ranks fifth in feature importance though it is still less important than $a$ by nearly an order of magnitude.

While useful, the feature importance in random forest classifiers should be interpreted cautiously, especially in cases where the features are correlated.
The time lags, as well as the maximum cross-correlations, in all channel pairs are very strongly correlated.
The emission measure slope is also likely correlated with the time lag and cross-correlation though perhaps more weakly so.
In particular, \citet{altmann_permutation_2010} found that as the number of correlated features in a random forest classifier increased, the individual importance of each feature in the correlated group decreased and that for a very large number of correlated features ($\sim50$), the feature importance of each was close to zero.
Here, we have at least two groups of 15 strongly correlated features each.
Thus, the values shown in \autoref{tab:importance} for the time lag and cross-correlation should be regarded as lower bounds on the feature importance. 
However, the presence of highly-correlated or unimportant features is not expected to affect the robustness or accuracy of the classifier.
 %
\section{Discussion}\label{sec:discussion}

As evidenced in \autoref{fig:probability-maps} and \autoref{fig:frequency-maps}, we find that high-frequency heating is likely to dominate in the core of the \AR. 
Comparing the heating frequency maps in \autoref{fig:frequency-maps} for the different cases in \autoref{tab:cases}, this high-frequency classification seems largely due to the steep observed emission measure slopes in the center of the \AR{} as seen in \autoref{fig:em-slopes}.
\added{Our heating classification is consistent with the results of \citet{warren_systematic_2012} who computed \dem{} for a single pixel near the periphery of \AR{} NOAA 1158 and found $a=2.7\pm0.32$.
This is consistent with our calculation of $a$ in this same region (see left panel of \autoref{fig:em-slopes}).
While \citet{warren_systematic_2012} cite this as evidence of ``high-frequency'' heating, we note that their definition of ``high-frequency'' overlaps with our classifcation of both high- and intermediate-frequency heating such that their conclusions are consistent with the classification results shown in \autoref{fig:frequency-maps}.}
Additionally, this result is consistent with X-ray observations of hot, steadier emission \citep{warren_evidence_2010,warren_constraints_2011,winebarger_using_2011} as well as the result of \citet{del_zanna_evolution_2015} who found high values of the emission measure slope in the center of NOAA 11193.

Comparing case C in \autoref{fig:frequency-maps} with the observed magnetogram of NOAA 1158 shown in Figure 1 of \citetalias{barnes_understanding_2019}, we find that areas of stronger magnetic field are spatially coincident with most of the pixels classified as high-frequency.
This suggests that those strands whose footpoints are rooted in areas of strong magnetic field strength are heated more frequently.
We will explore the relationship between the heating frequency and the underlying magnetic field strength in a future paper.

The longer loops surrounding the core are consistent with intermediate frequency heating.
Notably, the results from our classifier imply that low-frequency heating, as defined by \autoref{eq:heating_types}, is not needed to explain the observed time lags, suggesting that the waiting time on each strand in this \AR{} is likely to be on the order of or less than $\tau_\textup{cool}$.
This result is consistent with that of \citet{bradshaw_patterns_2016} who found that intermediate and high frequency nanoflares both produced time lags consistent with observations while their cooling experiment, similar to our low-frequency nanoflares, showed fundamental disagreements with the observed time-lag maps.

Unlike the high- and intermediate-frequency cases, we find that the groups of pixels most consistent with low-frequency heating do not appear as spatially-coherent loop-like structures, but instead have a ``patchy'' appearance.
This is especially true in cases C and D.
In particular, in the bottom row of \autoref{fig:frequency-maps}, these low-frequency patches appear to be near the footpoints of longer loops in the \AR{}.
We hypothesize that this lack of spatial coherence in the identification of low-frequency heating could be due to multiple overlapping structures, consistent with higher-frequency heating, along the LOS.
Additionally, these structures may also be undergoing both low- and higher-frequency heating during our selected 12 h observing window.

Additionally, we note that one could, in principle, model an entire active region with only steady heating and still reproduce the distribution of observed emission measure slopes.
For example, the observed shallow slopes on periphery could be consistent with steady 1 MK, 2 MK, and 3 MK loops all emitting along the LOS.
Similarly, the steep slopes near the inner core are consistent with only steady 3 MK loops along the LOS.
This is also the case with $T_{peak}$
However, it has been exhaustively shown that truly steady heating, in which the energy deposition is constant in time, is not consistent with observed time lags or cross-correlation values \citep[e.g.][]{viall_signatures_2016}.
Thus, we do not explicitly test a steady heating model here and note that even our high-frequency heating model produces variability in the observed emission.

After the emission measure slope, $a$, the next three most important features in the classification are the maximum cross correlations for the 211-193, 193-171, and 211-171 \AA{} channel pairs.
These three channels, 211 \AA{}, 193 \AA{}, and 171 \AA{}, peak sequentially in temperature at 1.8, 1.6, and 0.8 MK, respectively (see \autoref{fig:aia-response}), suggesting that the plasma dynamics in this temperature range, which are dominated by radiative cooling and draining \citep[e.g.][]{bradshaw_cooling_2005,bradshaw_cooling_2010,bradshaw_new_2010}, are coupled to, and indicative of, the frequency at which energy is deposited in the plasma and that thermal conduction has not erased all signatures of the heating.
A strand heated by low-frequency nanoflares will be allowed to cool well below 1 MK, producing a strong cross-correlation in these channel pairs, while a strand heated by high-frequency nanoflares will rarely be allowed to cool below the equilibrium temperature such that the cross-correlation, particularly in the 171 \AA{} channel pairs, is likely to be relatively low.
This cooling behavior is illustrated for a single strand in Figure 3 of \citetalias{barnes_understanding_2019}.
The relative importance of features from these three channels is again consistent with \citet{bradshaw_patterns_2016} who found that short, but non-zero 211-193 \AA{} time lags, in combination with the prevalence of zero 171-131 \AA{} time lags, are inconsistent with low-frequency nanoflares.

Interestingly, we find that there are no channel pairs, either for the time lag or maximum cross-correlation, that include the 94 \AA{} channel in the ten most important features as shown in \autoref{tab:importance}.    
Observed time lags \citep{viall_evidence_2012,viall_survey_2017} show a transition between being dominated by the hot 94 \AA{} emission in the inner core to cool 94 \AA{}  emission in the periphery as evidenced by time lags changing from positive to negative, respectively.
Two proposed explanations for this switchover are that either impulsive heating in the cores is more energetic or it is more frequent.
The inability of the 94 \AA{} pairs to effectively discriminate between heating frequencies, as measured by the feature importance, points to the switchover being dominated by the energy rather than the frequency.
This is also confirmed by the top row of Figure 8 of \citetalias{barnes_understanding_2019} which shows the positive-negative switchover between the inner core and the periphery for all heating frequencies.
However, we note that, compared to the other \AR s studied by \citet{warren_systematic_2012} and \citetalias{viall_survey_2017}, our chosen \AR{}, NOAA 1158, is relatively cool.
In \AR s dominated by more hot 94 \AA{} emission, the 94 \AA{} channel pairs may have a higher feature importance.

While the maximum cross-correlation in the 211-193 \AA{} channel pair (see bottom row of \autoref{fig:correlations}) is very high across the whole \AR{}, the 193-171 \AA{} and 211-171 \AA{} maps (as well as the other 171 \AA{} pairs except for 171-131 \AA{}) show a comparatively low cross-correlation.
Combined with the heating frequency maps in \autoref{fig:frequency-maps} which indicate that the center of the \AR{} is consistent with high-frequency heating, this suggests that many of the loops in the core are kept from cooling much below 0.9 MK.

An important caveat to this method for systematic comparison as we have applied it here is that the random forest classifier trained on the simulated emission measure slopes, peak temperatures, time lags, and maximum cross-correlations cannot provide any assessment of the accuracy of our model as described in \citetalias{barnes_understanding_2019}.
The classifier can only say, out of the provided classes (high-, intermediate-, or low-frequency), which type of heating \textit{best} describes the data.
However, given another method for assessing the heating frequency or perhaps some alternative forward-modeling approach, a random forest classifier could be used to compare these two methods.
In this way, machine learning also provides a promising strategy for reconciling different modeling approaches.
 %
\section{Conclusions and Summary}\label{sec:conclusions}


\added{In \citetalias{barnes_understanding_2019}, we carried out a series of numerical simulations to understand how the frequency of energy deposition is manifested in observable signatures in quiescent active regions.
By combining potential field extrapolations, efficient hydrodynamic modeling, and our novel and efficient forward modeling pipeline, we produced AIA images of \AR{} NOAA 1158 for all six EUV channels for $\approx8$ h of simulation time for high-, intermediate-, and low-frequency heating.
From these simulated intensities, we computed the emission measure slope and the time lag for all possible AIA channel pairs in each pixel of the \AR{} for all heating frequencies.
We found that the emission measure slope becomes increasingly shallow as heating frequency decreases, but as the heating frequency increases, the distribution of slopes peaks at higher values and becomes more broad.
Additionally, as the heating frequency decreased, the spatial distribution of time lags was increasingly determined by the distribution of loop lengths over the \AR{}.
Importantly, we also found that negative time lags in channel pairs where the second channel is 131 \AA{} provide a possible diagnostic for $\ge10$ MK plasma.}

In this paper, the second in our series on constraining nanoflare heating properties, we have used predicted diagnostics from \citetalias{barnes_understanding_2019} to systematically classify each pixel of \AR{} NOAA 1158 in terms of frequency of energy deposition.
In particular, we first collect 12 h of full-resolution SDO/AIA observations of NOAA 1158 in six EUV channels: 94, 131, 171, 193, 211, and 335 \AA.
We then co-align each image to a single time such that a given pixel in each image corresponds to approximately the same spatial coordinate and then crop the image to an area of $500\arcsec$-by-$500\arcsec$ centered on the \AR{}.

Next, we time-average the intensities of all six channels and use the method \citet{hannah_differential_2012} to compute the emission measure distribution in each pixel of the \AR{}.
We compute the peak temperature of the emission measure distribution, $T_{peak}$, as well as the emission measure slope, $a$, by fitting $\log_{10}\textup{EM}\sim a\log_{10}T$ over the temperature range $8\times10^5\,\textup{K}\le T < T_{peak}$. 
Additionally, we apply the time-lag analysis of \citet{viall_evidence_2012} to the full 12 h of observations of NOAA 1158 and compute the time lag, $\tau_{AB}$, and maximum cross-correlation, $\max\mathcal{C}_{AB}$, in each pixel of the \AR{} for all possible pairs of the six EUV channels, 15 in total.

Finally, we train a random forest classifier using the predicted emission measure slopes, peak temperatures, time lags, and cross-correlations for three different heating frequencies from \citetalias{barnes_understanding_2019}.
We then use our trained model to classify each observed pixel as consistent with either high-, intermediate-, or low-frequency heating (where the frequency is parameterized relative to the loop cooling time) and map the heating frequency across the entire \AR{}.

Our results can be summarized as follows:
\begin{enumerate}
    \item The distribution of observed emission measure slopes overlaps with the distributions of predicted emission measure slopes for high-, intermediate-, and low-frequency heating, suggesting a range of heating frequencies across the \AR{}.
    \item High-frequency heating dominates in the center of \AR{} and is coincident with loops whose footpoints are rooted in strong magnetic field.
    \item Intermediate-frequency heating is more likely in longer strands surrounding the center of the \AR{}. In most pixels, low-frequency heating, as defined in \autoref{eq:heating_types}, is not needed to explain the observed diagnostics.
    \item The emission measure slope is the strongest single-measure predictor of the heating frequency. Radiative cooling and draining around $1-2$ MK as manifested in the maximum cross-correlation also appears to be a strong indicator relative to the time lags. However, the feature importance as determined by the classifier should be interpreted carefully.
\end{enumerate}

We have demonstrated an efficient and powerful technique for constraining the heating frequency in active region cores and, more broadly, for systematically comparing models and observations.
While we have applied this technique for a particular set of heating parameters and a particular forward model of a single \AR{}, we emphasize that this approach for comparing models and observations is broadly applicable to any set of heating inputs and forward modeling technique. 
Given that the diagnostics here are known to vary with age \citep[e.g.][]{schmelz_cold_2012,del_zanna_evolution_2015} and from one \AR{} to the next \citep{warren_systematic_2012,viall_survey_2017}, the next step is to apply this methodology to a large sample of \AR s to place strong constraints on the frequency of energy deposition in the magnetically-closed corona.
 
\acknowledgments
This research makes use of \added{version 0.9.5 \citep{stuart_mumford_2018_2155946}} of sunpy, an open-source and free community-developed solar data analysis package written in Python \citep{the_sunpy_community_sunpy_2020}.
\added{We use v0.9.5 in this work to maintain consistency with the software environment used in \citetalias{barnes_understanding_2019} which was completed prior to the v1.0 release of sunpy.}
SJB and WTB were supported by the NSF through CAREER award AGS-1450230.
WTB was supported by NASA’s \textit{Hinode} program.
\textit{Hinode} is a Japanese mission developed and launched by ISAS/JAXA with NAOJ as a domestic partner and NASA and STFC (UK) as international partners.
It is operated by these agencies in cooperation with ESA and NSC (Norway).
The work of NMV was supported by the NASA Supporting Research program.
The complete source of this paper, including the data, code, and instructions for training the classification model, can be found at \href{https://github.com/rice-solar-physics/synthetic-observables-paper-observations}{github.com/rice-solar-physics/synthetic-observables-paper-observations}.

\facility{SDO(AIA)}

\software{
    astropy \citep[v3.1.0,][]{the_astropy_collaboration_astropy_2018,the_astropy_collaboration_2018_4080996},
	dask \citep[v1.0.0,][]{rocklin_dask:_2015},
	drms \citep[v0.5,][]{glogowski_drms_2019,kolja_glogowski_2019_2572850},
    matplotlib \citep[v3.0.2,][]{hunter_matplotlib_2007,thomas_a_caswell_2018_1482099},
	numpy \citep[v1.15.4,][]{harris_array_2020},
	PythonTeX \citep[v0.16,][]{poore_pythontex_2015},
    scikit-learn \citep[v0.20,][]{pedregosa_scikit-learn_2011,olivier_grisel_2019_2582066},
	seaborn \citep[v0.9.0,][]{michael_waskom_2018_1313201},
	scipy \citep[v1.1.0,][]{virtanen_scipy_2020, pauli_virtanen_2018_1241501},
	SolarSoftware \citep{freeland_data_1998},
    sunpy \citep[v0.9.5,][]{stuart_mumford_2018_2155946}
}

\bibliographystyle{aasjournal.bst}
\bibliography{references.bib,software.bib}

\listofchanges

\end{document}